\documentclass[12pt]{iopart}
\usepackage{graphicx}
\usepackage{amstext}
\usepackage{bm}

\def\ml{\hbox to 0.2pt{$\langle$}\hbox to 0.2pt{$\langle$}\hbox to 0.2pt{$\langle$}\langle}
\def\mr{\hbox to 0.2pt{$\rangle$}\hbox to 0.2pt{$\rangle$}\hbox to 0.2pt{$\rangle$}\rangle}

\def\be{\begin{equation}}
\def\ba{\begin{eqnarray}}
\def\ee#1{\label{#1}\end{equation}}
\def\ea#1{\label{#1}\end{eqnarray}}
\def\la{\langle}
\def\ra{\rangle}
\def\bs{\begin{center}}
\def\es{\end{center}}

\graphicspath{{img/}}

\begin{document}

\title{Optimal strategy for controlling transport in inertial Brownian motors}

\author{
Lukasz Machura$^{1,2}$, Marcin Kostur$^1$, Fabio Marchesoni$^{3}$,
Peter Talkner$^1$, Peter H\"anggi$^1$, Jerzy {\L}uczka$^2$ }

\address{$^1$\ Institute of Physics, University of Augsburg,
Universit\"atsstrasse 1, D-86135 Augsburg, Germany}
\address{$^2$\ Institute of Physics, University of Silesia,
P-40-007 Katowice, Poland}
\address{$^3$\ Dipartimento di Fisica,
Universit\`{a} di Camerino, I-62032 Camerino,  Italy}

\ead{lukasz.machura@physik.uni-augsburg.de}


\begin{abstract}
In order to optimize the directed motion  of an
inertial Brownian motor, we identify the operating conditions that both maximize
the motor current and minimize its dispersion. Extensive numerical simulation
of an inertial rocked ratchet displays that two quantifiers, namely
the energetic efficiency and the P\'eclet number (or equivalently
the Fano factor), suffice to determine the regimes of optimal transport.
The effective diffusion of this rocked inertial Brownian motor can be expressed 
as a generalized fluctuation theorem of the Green -- Kubo type. 
\end{abstract}
\submitto{\JPC} 
\maketitle


\section{Introduction}
The theoretical concepts of Brownian motors and ratchet transport
\cite{BM} have been experimentally
 realized in a
variety of systems.  Examples are: cold atoms in optical lattices
\cite{jones04}, colloidal particles in holographic optical trapping
patterns \cite{lee05}, ratchet cellular automata \cite{babic05},
superconducting films with  periodic arrays of asymmetric pinning
sites \cite{vondel05,villegas05}, to mention only a few.

When we study  the motion of Brownian motors, the natural transport
measure is a  conveniently defined average asymptotic velocity $\la
v \ra$ of the Brownian motors. It describes how much time  the
typical particle needs to overcome  a given distance in the
asymptotic (long-time) regime. This velocity, however,  is not the
only relevant transport criterion. Other attributes can also be
important. In order to establish these, we consider  the two
following aspects: the quality of the transport and the energetic
efficiency of such a system.
\begin{figure}[htbp]
\bs
\includegraphics[scale=1]{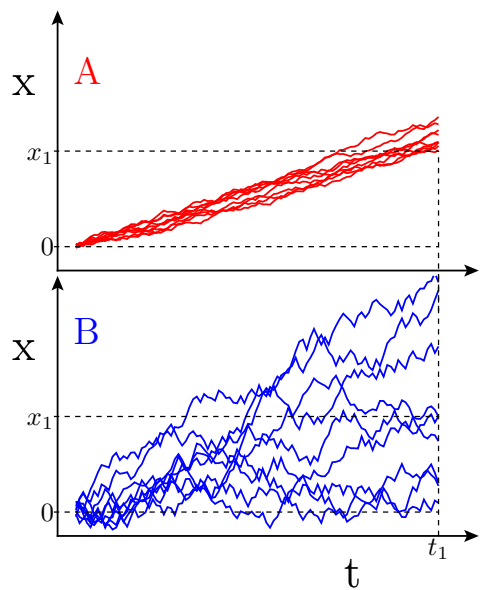}
\caption{(Color online) Two sets of illustrative trajectories of an
inertial, rocked Brownian motor (see in text). Both sets of
trajectories A and B possess the same average asymptotic velocity,
but exhibit a distinct different diffusion behavior.} \label{mmfig1}
\es
\end{figure}

In  Fig. \ref{mmfig1}, one can identify two different groups A and B
of  random trajectories of the Brownian particle; both possess the
same average drift velocity $\la v \ra$. However, it is obvious upon
inspection that the dynamical properties of these two groups of
trajectories are different. The particles from the group A travel
more or less coherently together while the particles from the group
B spread out as  time goes by. If we fix the distance $x=x_1$ then
most particles from the group A reach this distance at about the
same time $t=t_1$, while for $t=t_1$ most of the B trajectories
either stay behind or have already proceeded to more distant
positions. It is thus evident that the noise-assisted, directed
transport for the particles in the group A is more effective than in
the group B.

\begin{figure}[htbp]
\bs
\includegraphics[scale=1]{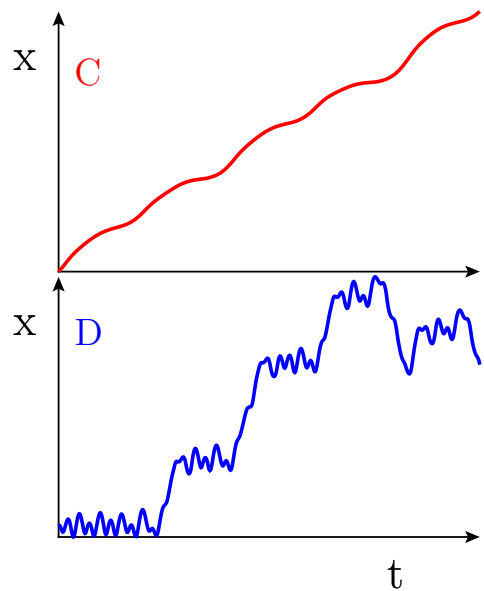}
\caption{(Color online) Typical trajectories of an inertial,
rocking Brownian motor; both sets assume the same average velocity
but differing velocity fluctuations.} \label{mmfig2} \es
\end{figure}

There is still another efficiency aspect related to Brownian motor transport. This
refers to the external energy input into the system which may be
essential in practical applications. We like to know how much of
this input energy
 is converted into useful work, namely into directed cargo transport, and
how much of it gets wasted. Since motors move in a dissipative
environment, we need to know how much of the input energy is being
spent for moving a certain distance against the acting friction
force. Fig. \ref{mmfig2} depicts trajectories representing different
motor scenarios. The motor C moves forward unidirectionally, while
the motor D moves in a  more complicated manner: its motion
alternates small oscillations and fast episodes, mostly in the
forward direction,  but sometimes also in the backward direction.
Again the mean velocity in both cases is the same, however,  the
particle C uses energy pumped from the environment to proceed
constantly forward while the particle D wastes part of its energy to
perform oscillations and back-turns. By simply inspecting these
schematic pictures one can guess immediately when directed transport
is more effective.

We note  that in Fig. \ref{mmfig1},  the cases A and B can be
characterized by the effective diffusion coefficient $D_{eff}$,
i.e.,  by the spreading of fluctuations in the position space while
the cases C and D in Fig. 2 can be characterized by
 the variance of velocity  $\sigma_v^2 = \la v^2 \ra - \la v \ra^2$.
The three quantities  $\la v \ra$,
$D_{eff}$ and $\sigma_v^2$ can be combined to define two important
characteristics of transport, namely the efficiency of noise
rectification and the so-called P\'eclet number \cite{peclet}.

Our work  is organized as follows. In the following section, we
detail the model of an  inertial rocked Brownian motor. In section
3, we present a general discussion of the efficiency measures
 of  Brownian motors.
In section 4, a description of the ratchet based on point processes
is introduced. In sections 5 and 6, our numerical findings are
analyzed in the context of the optimization conditions for transport of
inertial Brownian motors.  A summary is provided in section 7.


\section{Inertial  rocked Brownian motors}

The archetype of the inertial Brownian motor is
represented by a classical particle of mass $m$
moving in a spatially periodic and  asymmetric  potential
$V(x)=V(x+L)$ with period $L$ and barrier height $\Delta V$
\cite{jung96,lindner99}.  The particle is driven by an external,
unbiased, time-periodic force of amplitude $A$ and angular
frequency $\Omega$ (or period $
 {\cal T}_{0} = 2\pi / \Omega$). The system is additionally subjected to
thermal noise $\xi(t)$.
\begin{figure}[htbp]
\bs
\includegraphics[scale=0.9]{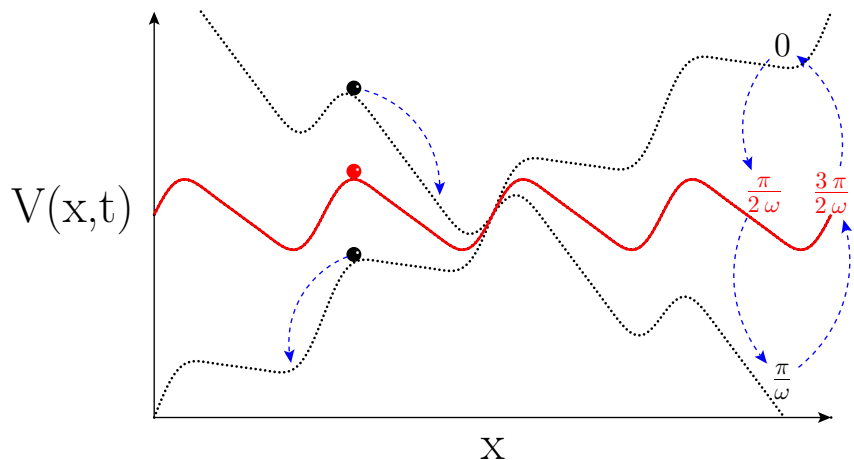}
\caption{(Color online) Schematic picture of a rocking ratchet with
the potential  $V(x,t) = V(x)-x a \cos(\omega t)$, cf. Eqs (2) and (4).}
\label{mmfig3}
\es
\end{figure}
The  dynamics of the system is modeled by
the Langevin equation
\cite{hanggi1982}
\begin{eqnarray}
\label{eq:Lan1} m \ddot x + \gamma \dot x = -V'(x) + A \cos(\Omega
t) + \sqrt{2\gamma k_B T} \; \xi(t),
\end{eqnarray}
where a dot denotes differentiation with respect to time and a
prime denotes a differentiation with respect to  the Brownian motor
coordinate $x$.
 The parameter $\gamma$ denotes the Stokes friction coefficient,
$k_B$ the Boltzmann constant and  $T$ is the temperature. The
thermal fluctuations due to the coupling of the particle with the
environment are modeled by a zero-mean, Gaussian white noise
$\xi(t)$ with auto-correlation function $\langle \xi(t)\xi(s)\rangle
= \delta(t-s)$ satisfying Einstein's fluctuation-dissipation
relation.

Upon introducing characteristic length scale and time scale, 
Eq. (\ref{eq:Lan1}) can be rewritten in dimensionless form, namely
\begin{equation}
\ddot{\hat{x}} + \hat{\gamma} \dot{\hat x} =- \hat{V}'(\hat{x}) + a
\cos(\omega \hat{t}) + \sqrt{2\hat{\gamma}D_0} \; \hat{\xi}
(\hat{t}), \label{NLbw}
\end{equation}
with \cite{mach}
\begin{eqnarray}
\hat{x} = \frac{x}{L}, \qquad \hat{t} = \frac{t} {\tau_0},
 \qquad \tau_0^2 = \frac{mL^2}{\Delta V}.
\end{eqnarray}
The characteristic time $\tau_0$ is  the time a particle of mass $m$
needs to move a distance $L/2$ under the influence of the constant
force $\Delta V/L$ when starting with velocity zero. The remaining
re-scaled parameters are:
\begin{itemize}

\item the friction coefficient $\hat{\gamma} = (\gamma / m)
\tau_0 = \tau_0 / \tau_L$ is the ratio of the two characteristic
times, $\tau_0$ and the relaxation time  of the velocity degree of
freedom, i.e., $\tau_L = m/\gamma$,
\item  the potential
$\hat{V}(\hat{x})=V(x)/\Delta V = \hat{V}(\hat{x}+1)$ has unit
period and unit  barrier height $\Delta \hat{V}=1$,
\item the amplitude $a = A L / \Delta V$
and the  frequency $\omega = \Omega \tau_0$
(or the period ${\cal T}=2\pi/\omega$),
\item  the zero-mean white noise
$\hat{\xi}(\hat{t})$ has auto-correlation function
$\langle\hat{\xi}(\hat{t})\hat{\xi}(\hat{s})\rangle=\delta(\hat{t}-\hat{s})$
with re-scaled noise intensity $D_0 = k_B T / \Delta V$.
\end{itemize}
From now on,  we will  use only the dimensionless variables and
shall omit the ``hat'' for all quantities in Eq. (\ref{NLbw}).

For the asymmetric ratchet potential $V(x)$ we consider a linear
superposition of three spatial harmonics \cite{mach},
\begin{eqnarray}\label{pot}
V(x) = V_0 [\sin(2 \pi x) + c_1 \sin (4 \pi x) + c_2 \sin (6 \pi
x)],
\end{eqnarray}
where $V_0$ normalizes the barrier height to unity and the
parameters $c_1$ and $c_2$ determine the ratchet profile. Below, we
analyze the case when $c_1=0.245$ and $c_2=0.04$. Then $V_0 =0.461$.
This potential is shown as a bold (red) line in Fig. \ref{mmfig3}.


\section{Quantifiers characterizing optimal transport of Brownian motors}

As already elucidated above, there are several quantities that
characterize the effectiveness of directed transport. The effective
diffusion coefficient, describing the fluctuations around the
average position of the particles, is defined as
\be D_{eff} = \lim_{t \rightarrow \infty} \frac{\la x^2(t) \ra - \la
x(t) \ra^2}{2t}, \ee{Deff}
where the brackets $\la \dots \ra$ denote an average over the
initial conditions of position and velocity and over all
realizations of the thermal noise. The coefficient $D_{eff}$ can also
be introduced via a generalized Green-Kubo relation which we detail in
the Appendix. Intuitively, if the stationary velocity is large and
the spread of trajectories is small,  the diffusion coefficient is
small and the transport is more effective. To quantify this, we can
introduce the dimensionless P\'eclet number Pe \cite{peclet,landau}
by use of a double-averaging procedure, i.e.,
\ba \mbox{Pe} = \frac{ L \la\la v \ra\ra }{D_{eff}}, \ea{factors}
where  the "double-average" $\la \!\la v \ra\!\ra$ denotes the
average of the asymptotic velocity over one cycle of the external
drive, i.e.,
\ba \la\!\la  v \ra\!\ra = \lim_{t \to \infty} \frac{1}{t}
\int_0^{t} \la v(t')\ra\; dt' = \frac{\omega}{2\pi}
\int_0^{2\pi/\omega} \la v(t')\ra_{as}\; dt', \ea{aver}
where the average $\la \dots \ra_{as}$ in the second integral refers
to the asymptotic periodic state.

Originally, the P\'eclet number Pe  arises in problems of heat
transfer in  fluids where it stands for the ratio of heat advection
to diffusion. When the P\'eclet number is small, the random motion
dominates; when it is large, the ordered and regular
 motion dominates.
The value of the P\'eclet number depends on some characteristic length scale of the
system. Dealing with ratchets the most adequate choice for such length scale is the period
of the periodic potential, which in re-scaled units is equal to $1$.

The second aspect of the motor trajectories we want to control has to do with the
fluctuations of the velocity $v(t)$.  In the long-time  regime, it is characterized by
the variance $\sigma_v^2 = \la\!\la v^2 \ra\!\ra - \la\!\la v
\ra\!\ra^2$. The Brownian motor moves with an actual velocity
$v(t)$, which is typically contained within the interval
\begin{eqnarray}
 \label{v(t)}
v(t) \in \left( \la\!\la v \ra\!\ra -\sigma_v, \la\!\la v \ra\!\ra
+\sigma_v\right).
\end{eqnarray}
Now, if $\sigma_v > \langle \!\la v \ra\!\rangle$,  the Brownian
motor may possibly move for some time in the direction opposite to
its average velocity $\langle \!\la v \ra\!\rangle$ and the directed
transport becomes less efficient. If we want to optimize the
effectiveness of the motor motion we must introduce a measure for
the efficiency $\eta$ that accounts for the velocity fluctuations,
too, namely \cite{munakata}
\begin{eqnarray}
\label{eta} \eta  =  \frac{ \la\!\la v \ra\!\ra^2} {|\la\!\la v
\ra\!\ra^2 +\sigma_v^2 - D_0|} =  \frac{\la\!\la v \ra\!\ra^2}
{|\la\!\la v^2 \ra\!\ra - D_0|}\;.
\end{eqnarray}
This definition follows from an energy balance of the underlying
 equation of motion  (\ref{NLbw})
(see the Appendix of Ref. \cite{mach}).
 If  the variance of velocity $\sigma_v$ is reduced,
the energetic efficiency (\ref{eta}) increases and the transport of
the Brownian motor becomes  more efficient.

\section{A corresponding point process related to the rocked Brownian motor dynamics}

\begin{figure}[htbp]
\bs
\includegraphics[scale=0.7,clip=]{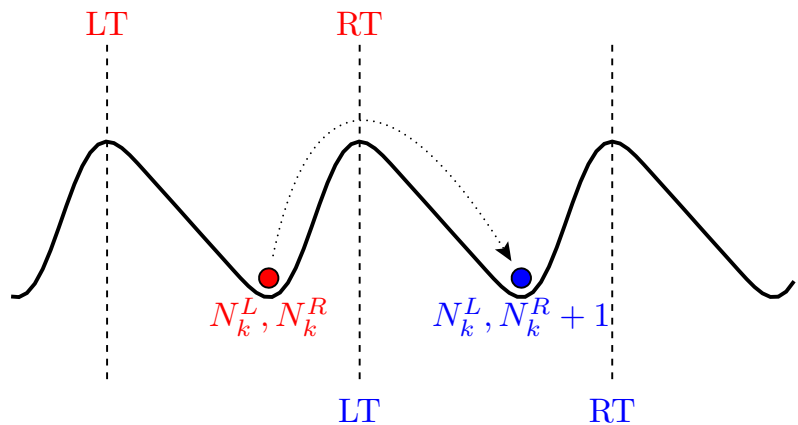}
\caption{(Color online) The point process related to rocked Brownian
motor dynamics. Two thresholds LT and RT  are the corresponding
maxima, to the left and to the right of the  particle position. If
the particle jumps over the right neighboring barrier, the
 number $N_k^{R}$ increases to $N_k^R +1$ and the previous
right sided threshold assumes the role of the left sided threshold.
}
 \label{mmfig4} \es
\end{figure}

The running trajectories can be characterized in a coarse grained
way by only counting the events when a trajectory traverses from
one potential well into a neighboring one, and by disregarding the
details of the intra-well motion, see figure \ref{mmfig4}. In this way, a point process can
be introduced that can be investigated in a standard way
\cite{talkner-rice}. Most, though not all, of the quantities
describing the original continuous process can be retrieved from the
so defined point process.
For  this purpose we introduce two random, natural numbers
$N_k^{\alpha}$, where $\alpha = \{R, L\} $ stands for right (R) and
left (L). The number $N_k^{R}$ is given by the number of barrier
crossings towards the right within the $k$-th period of the driving,
i.e. in the time between $(k-1){\cal T}$ and  $k{\cal T}$. The
respective number of barrier crossings to the left is denoted by
$N_k^L$. The difference
\be N_k = N_k^R - N_k^L \ee{N}
 indicates that  during a temporal period $\cal T$ the particle has covered
the distance $x_k = N_k L = N_k$. Hence the average, asymptotic
velocity is given by
\begin{eqnarray}  \label{v}
\la\!\la v \ra\!\ra =\lim_{t \to \infty} \frac{1}{t} \int_0^{t} \la
v(t')\ra\; dt' =\lim_{K \to \infty} \frac{1}{K{\cal T}}\sum_{k=1}^K
\;
 \int_{(k-1){\cal T}}^{k{\cal T}} \la v(t')\ra\; dt'\nonumber\\
= \lim_{K \to \infty}  \frac{1}{K{\cal T}}\sum_{k=1}^K x_k = \lim_{K
\to \infty}  \frac{1}{{\cal T} K}\sum_{k=1}^K N_k = \frac{ \la N
\ra}{{\cal T}}.
\end{eqnarray}
Analogously, the effective diffusion coefficient is determined by
the relation
\be D_{eff} =  {\la \delta N^2 \ra \over 2 {\cal T}} = \; {\la N^2
\ra - \la N \ra^2 \over 2 {\cal T}}. \ee{deff}
A related quantity is the Fano factor
$F$ \cite{fano}, defined here as
the fluctuation to the first moment ratio
\be F = {\la \delta N^2 \ra \over \la N \ra} . \ee{FF}
As such, the Fano factor provides a quantitative measure of the
relative number fluctuations or the relative randomness of the
process; in the case of a Poisson
process $F = 1$.

On the other hand, from (\ref{factors}), (\ref{v}) and (\ref{deff})
it follows that the P\'eclet number can be expressed as
\be \mbox{ Pe} = {2 \la N \ra \over \la \delta N^2 \ra}. \ee{Pe}
This quantifier is thus related to the Fano factor via the relation
$\mbox{Pe} = 2/F$.

\section{Numerical analysis}
The noiseless, deterministic inertial rocked ratchet shows a rather
complex behavior and, in distinct contrast to  overdamped rocked
Brownian motors \cite{ODBM}, often exhibits a chaotic dynamics
\cite{jung96,chaotic}. By adding noise, one typically activates a
diffusive dynamics; thus allowing for stochastic escape events among
possibly coexisting attractors. As analytical mathods to handle these 
situations  effectively do not exist, we carried out
extensive numerical simulations. We have numerically integrated Eq.
(\ref{NLbw}) by the Euler method with time step $h=5 \times
10^{-4}\;{\cal T}$. The initial conditions for the coordinate $x(t)$
were chosen according to a uniform distribution within one cell of
the ratchet potential. The starting velocities of the particles were
also distributed uniformly in the interval $[-0.2,0.2]$.

The first $10^3$ periods $\cal T$ of the external force were skipped
in order to avoid transient effects. We
employed two tactics of extracting the above characteristics from
the generated trajectories. For the estimation of the energetic
efficiency (or velocity fluctuations) the usual averages over the
time ($10^5\;{\cal T}$) and $333$ different realizations were
taken. In the the  case of the P\'eclet number (or effective
diffusion)  we  used the point process approach (see previous
section for details); therefore, only one time average of
long-time runs ($10^6\;{\cal T}$) was required.

\begin{figure}[htbp]
\bs
\includegraphics[scale=0.8]{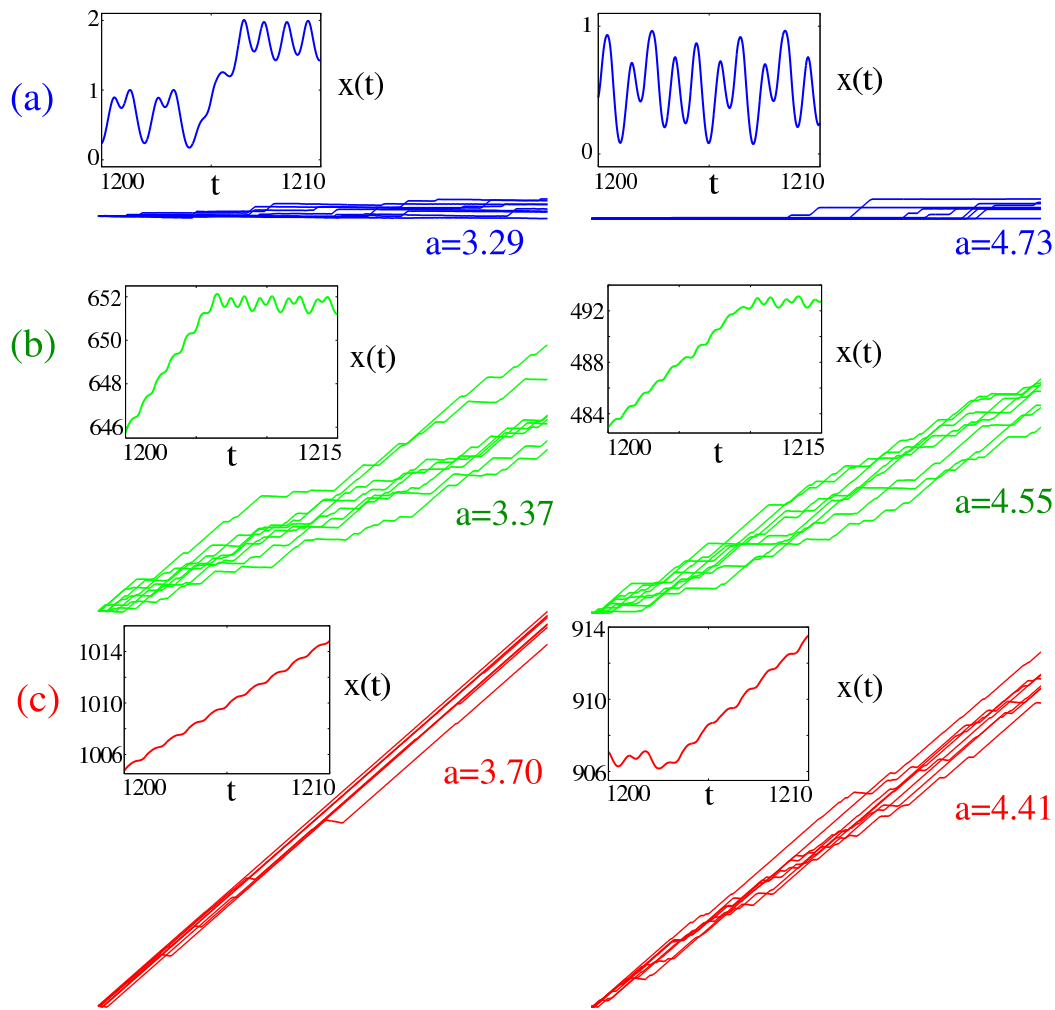}
\caption{(Color online) Brownian trajectories of the rocked particle
moving in the asymmetric ratchet potential $V(x) = V_0 [\sin(2 \pi
x) + 0.245
  \sin (4 \pi x) + 0.04 \sin (6 \pi x)]$, where $V_0\simeq 0.461$
  normalizes the barrier height to unity.  The
  forces stemming from such a potential range between $-4.67$ and
  $1.83$. The two angular frequencies  at the well-bottom and at the
  barrier-top are  the same, reading    $5.34$.
The remaining parameters are: $\gamma=0.9$, $\omega = 4.9$ and
$D_0=0.001$.  The values of the  driving amplitude are $a=3.29,
3.37, 3.70, 4.41, 4.55, 4.73$. One can see that for  $a=3.29$ and
$4.73$ in part (a) (in blue online), the particles usually oscillate in a
potential well, most of the time performing only a few steps.
This results in  an
almost zero mean velocity, a very small effective diffusion but with
rather large velocity fluctuations. For another set of driving
amplitudes: $a=3.37$ and $4.55$ in part (b) (in green online) the mean
velocity is large, $\sigma_v$ becomes suppressed, but the effective
diffusion exhibits an enlargement due to a 'battle between
attractors'. Part (c): The cases $a=3.70$ and $a=4.41$ (in red online)
correspond to the optimal {\it modus operandi} of the inertial
Brownian motor - the net drift is maximal and fluctuations
get suppressed.} \label{mmfig5} \es
\end{figure}


\begin{figure}[htbp]
\bs
\includegraphics[clip=]{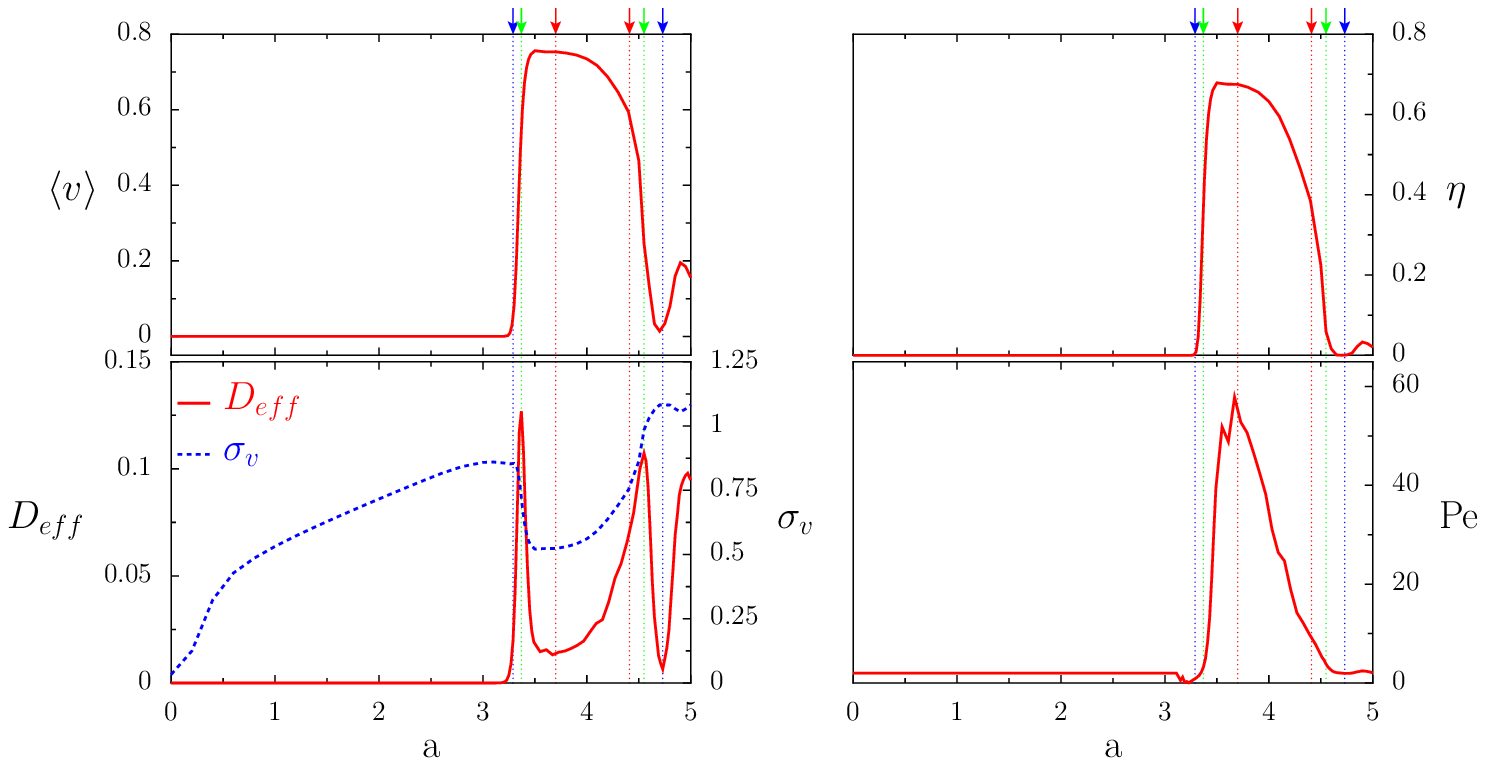}
\caption{(Color online) Left top: the average, dimensionless
  velocity $\langle v\rangle$ of the inertial, rocked Brownian motor under
  nonadiabatic driving conditions. Left bottom: corresponding velocity
  fluctuations $\sigma_v$ (dotted line) and corresponding diffusion coefficient $D_{eff}$
  (solid line). Right top: Brownian motor efficiency $\eta$. Right bottom: depicted is the P\'eclet number Pe,
  being proportional the  inverse of the Fano factor.  All
  quantities are plotted versus the external driving amplitude $a$.
Values of the remaining parameters are the same as in
Fig.~\ref{mmfig5}; e.g. the thermal noise strength  here is $D_0 =
0.001$. The numerical errors are within the line width. }
\label{mmfig6} \es
\end{figure}

Typically, there are two possible dynamical states of the ratchet system: a
locked state, in which the particle oscillates mostly within one
potential well (cf. the case with amplitude a = 4.73 in Fig.
\ref{mmfig5}), and a running state, in which the particles surmount
the  barriers of the potential. Moreover, one can distinguish  two
classes of running states: either the  particle overcomes the
barriers without  any back-turns (cf. the case with an amplitude a =
3.70 in Fig.  \ref{mmfig5}) or it undergoes frequent oscillations
and back-scattering events (cf. the case with amplitude a = 4.55 in
Fig.  \ref{mmfig5}). For a small driving amplitude, we find that the
locked behavior is  generic implying that the average motor velocity
is almost zero, see Fig. \ref{mmfig5}. If the amplitude is increased
up to some critical value, here $a=3.25$, the running solutions
emerge. Around that critical point, there occurs a 'battle of
attractors' and the particle burns
 energy for both barrier crossings and intra-well oscillations.
This behavior is reflected in  an enormous enhancement of the
effective diffusion \cite{giantdiff}.

If  the driving amplitude is further increased, a regime  of optimal
transport sets in. The rapid growth of the average
velocity is accompanied by a decline of both the position and the
velocity  fluctuations. It means that the trajectories bundle closely
together; note the case  $a=3.70$ in  Fig.
\ref{mmfig5}. Because there are no intra-well oscillations, the
energy that gets dissipated per unit distance, is minimal.

At  even larger drive amplitudes an upper threshold
 is approached (in the present case this threshold is located at around $a=4.7$)
where the velocity sharply  decreases  to a value close to zero.
Moreover,  the diffusion coefficient is small and  the velocity
fluctuations are large, cf. the case with amplitude $a=4.73$ in Fig.
\ref{mmfig5}. In this regime, the particle dangles around
its actual  position, as it occurs for $a < 3$, meaning that its
motion is confined mostly to one well. We note, however, that the
amplitude of the intra-well oscillations becomes much larger so that
the corresponding velocity fluctuations are also large.

We conclude that the diffusion coefficient is small for cases when
the particle performs either locked motion or running motion without
back-turns.

All these considerations are accurately encoded and described by the
two previously discussed measures, namely,   the efficiency
(\ref{eta}) and the P\'eclet number Pe in (\ref{factors}) or in
(\ref{Pe}). It is found  that the optimal regime for the ideal modus
operandi  of the Brownian motor is achieved when both the efficiency
and the P\'eclet number become maximal, see in Fig. \ref{mmfig6}. Indeed,
in this regime of optimal performance, the particle moves forward
steadily, undergoing rare back-turns \cite{porto}, see the
case $a= 3.70$ in Fig. \ref{mmfig6}.


\section{Role of temperature}

\begin{figure}[htbp]
\bs
\includegraphics[clip=]{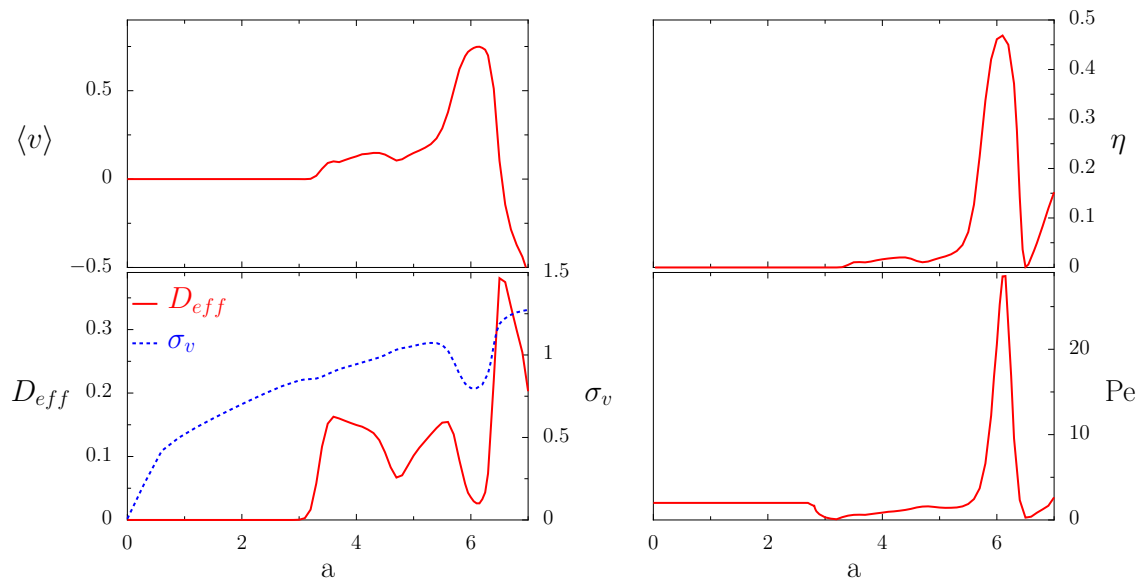}
\caption{(Color online) The same as in Fig. \ref{mmfig6}  but here
for a five times larger  temperature, i.e.
 $D_0 = 0.005$.}
\label{mmfig7} \es
\end{figure}

We next address the dependence on the strength of thermal noise. In
Fig. \ref{mmfig7}, we present our numerical results for the
noise-assisted, directed transport at a larger temperature, namely
for  $D_0 = 0.005$. The potential barrier height is
still rather high in comparison to the thermal energy. In this
regime, a so-called current reversal, i.e. a change  of the
transport direction occurs as a function of the driving amplitude.
Otherwise, the behavior remains qualitatively the same as for
lower noise  $D=0.001$. The diffusion coefficient
exhibits three maxima and two minima in the corresponding interval
of the drive amplitudes. However, the optimal regime corresponds to
the neighborhood of the second minimum of the diffusion coefficient.
In contrast, we notice that at lower noise $D=0.001$,
the optimal regime set in within the neighborhood of the first
minimum of $D_{eff}$.

Under higher temperature
operating conditions, optimal transport also occurs when both
the efficiency $\eta$ and the P\'eclet number $Pe$ are maximal.

\section{Summary}

In this work, criteria for the  optimal transport of an inertial
rocked ratchet were  established using two characteristic
quantifiers: the energetic efficiency (\ref{eta}) and the P\'eclet
number (\ref{factors}). Adapting the methods of point processes to
rocked Brownian motors, we expressed the averaged motor velocity and
the position-diffusion coefficient by corresponding averages of the
point process $N_k$. Both these measures can be  obtained from
simulations of the driven Langevin dynamics (\ref{NLbw}).

The Fano factor $F$ used in the theory of point processes is related
to the P\'eclet number in a simple manner via Pe$=2/F$. In our case,
it is more convenient to  employ the P\'eclet number because in
regimes where the average velocity is very small, the P\'eclet
number assumes  values close to zero, while the Fano factor would
diverge. From our numerical analysis it follows that the optimal
modus operandi for the inertial Brownian motor is obtained when the
efficiency $\eta$ and the P\'eclet  number assume simultaneously
maximal values.

\section*{Appendix}

In the present paper we have considered  the effective diffusion
coefficient, which is defined as
\be D_{eff} = \lim_{t \rightarrow \infty} \frac{\la x^2(t) \ra - \la
x(t) \ra^2}{2t}, \ee{Deg}
where the brackets $\la \dots \ra$ denote an average over the
initial conditions of position and velocity and over all
realizations of the thermal noise. Another definition of the
diffusion coefficient is given by the formula
\be D = \lim_{t \rightarrow \infty} \frac{\la [\delta x(t)
 -  \delta x(0)]^2 \ra}{2t},
\ee{D2}
where $\delta x(t) = x(t) - \la x(t)\ra$. By inspection one finds
\be
 D_{eff} = D
\ee{Dif}
if
\be
 \lim_{t \rightarrow \infty} \frac{1}{t} \la \delta x(t)
  \delta x(0) \ra =0.
\ee{delt}
In our case, this term vanishes  because of the presence of thermal
noise and dissipation. More generally, $| \la  \delta x(t) \delta
x(0) \ra |$ may increase at most as $t^{1/2}$ if the diffusion
coefficient $D$ as defined in (\ref{D2}) is finite. Consequently,
for
 such  processes the equation (\ref{Dif}) also holds.

We now show  that  the diffusion constant $D$ is related to the
auto-correlation function of the velocity via a Green-Kubo relation,
in spite of the fact that the system is far from equilibrium. For a
system with periodic driving, $D$ takes the form
\be
 D =  \int_0^{\infty} ds \; \overline C(s),
\ee{kubo}
where
\ba {\overline C(s)} = \frac{1}{\cal T} \int_0^{\cal T} d\tau \;
C_{as}(\tau,s) \ea{veve}
denotes the time average of the velocity correlation function
$C_{as}(\tau, s)$ over  one period ${\cal T}=2\pi/\omega$
of the driving and where
\ba C_{as}(t, s)=  \la\delta v(t) \delta v(t+s) \ra_{as} = \la\delta
v(t+s) \delta v(t) \ra_{as} \ea{vev}
is the nonequilibrium asymptotic velocity-velocity correlation
function. In the case of  periodic driving, this function is
periodic with respect to the first argument, i.e.,
\ba
 C(t,s) = C(t+{\cal T}, s).
\ea{peri}
To show the Green-Kubo relation, we start from the expression
 $\dot x(t)=v(t)$ from which it follows that
\be \delta x(t) - \delta x(0)   =  \int_0^{t} ds \; \delta v(s).
\ee{vel}
Therefore (\ref{D2}) takes the form
\begin{eqnarray} \label{DD}
2 D = \lim_{t \rightarrow \infty} \frac{1}{t} \int_0^{t} ds_1
\int_0^{t} ds_2 \;
\la\delta v(s_1) \delta v(s_2)\ra \nonumber\\
= \lim_{t \rightarrow \infty} \frac{1}{t} \int_0^{t} ds_1 \int_0^{t}
ds_2 \; C(s_2, s_1-s_2),
\end{eqnarray}
where \ba C(t, s)=  \la\delta v(t) \delta v(t+s) \ra. \ea{cik}
Changing the integration variables $(s_1, s_2)\to(s = s_1-s_2,\
\tau=s_2)$ and exploiting  the symmetry of the correlation function,
$ C(t, s) = C(t+s, -s)$,  one obtains
\begin{eqnarray} \label{Dd}
 D = \lim_{t \rightarrow \infty} \frac{1}{t}
\int_0^{t} ds \int_0^{t-s} d\tau \; C(\tau, s) \nonumber\\
 =\lim_{t \rightarrow \infty} \frac{1}{t}
\int_0^{t} ds \int_0^{t} d\tau \; C(\tau, s) - \lim_{t \rightarrow
\infty} \frac{1}{t} \int_{0}^{t} ds \int_{t-s}^{t} d\tau \; C(\tau,
s).
\end{eqnarray}
We assume that the diffusion coefficient is finite. Therefore
 the second term in the second line of (\ref{Dd})
tends to zero as $t\to\infty$, so that
\begin{eqnarray} \label{Fs}
D = \int_0^{\infty} ds \; \lim_{t\to\infty}  \frac{1}{t}
 \int_0^{t} d\tau\; C(\tau,s).
\end{eqnarray}
For $t=K{\cal T}$, one splits the second integral into sum over
subsequent periods,
\begin{eqnarray} \label{Fi}
 \lim_{t\to\infty}  \frac{1}{t}
 \int_0^{t} d\tau\; C(\tau,s)
= \lim_{K\to\infty}  \frac{1}{K{\cal T}}\sum_{k=1}^K \;
 \int_{(k-1){\cal T}}^{k{\cal T}} d\tau \; C(\tau,s ) \nonumber\\
= \frac{1}{\cal T}  \int_0^{\cal T} d\tau \; C_{as}(\tau, s) =
{\overline C(s)}
\end{eqnarray}
where
\begin{eqnarray} \label{Fij}
 C_{as}(\tau, s)= \lim_{K\to\infty}\frac{1}{K}
\sum_{k=0}^K \; C(\tau+k{\cal T},s).
\end{eqnarray}
The Eqs. (\ref{cik}), (\ref{Fs})-(\ref{Fij}) represent the
Green-Kubo relation for the diffusion constant of such periodically
driven processes  $x(t)$; notably, these {\it per se} constitute far
from equilibrium  processes.

\ack The authors gratefully acknowledge financial support by the
Deutsche Forschungsgemeinschaft via grant HA 1517/13-4, the
Graduiertenkolleg 283 (LM, PT, PH), the collaborative research grant
SFB 486, the DAAD-KBN (German-Polish project {\it Stochastic
Complexity}) (PH and J{\L}), and the ESF, Program {\it
Stochastic Dynamics: fundamentals and applications; STOCHDYN} (PH and FM).

\section*{}
\renewcommand{\theequation}{E\arabic{equation}}
\setcounter{equation}{0}
\newpage
\setcounter{page}{1}
\begin{center}
  {\bf ADDENDUM AND ERRATUM}
\end{center}
\vspace*{5mm}
\title{Optimal strategy for controlling transport in inertial
  Brownian motors} 
  
\author{
Lukasz Machura$^{1,2}$, Marcin Kostur$^1$, Fabio Marchesoni$^{3}$,
Peter Talkner$^1$, Peter H\"anggi$^1$, Jerzy {\L}uczka$^2$ }

\address{$^1$\ Institute of Physics, University of Augsburg,
Universit\"atsstrasse 1, D-86135 Augsburg, Germany}
\address{$^2$\ Institute of Physics, University of Silesia,
P-40-007 Katowice, Poland}
\address{$^3$\ Dipartimento di Fisica,
Universit\`{a} di Camerino, I-62032 Camerino,  Italy}

\begin{figure}[b]
\begin{center}
\includegraphics[scale=0.7]{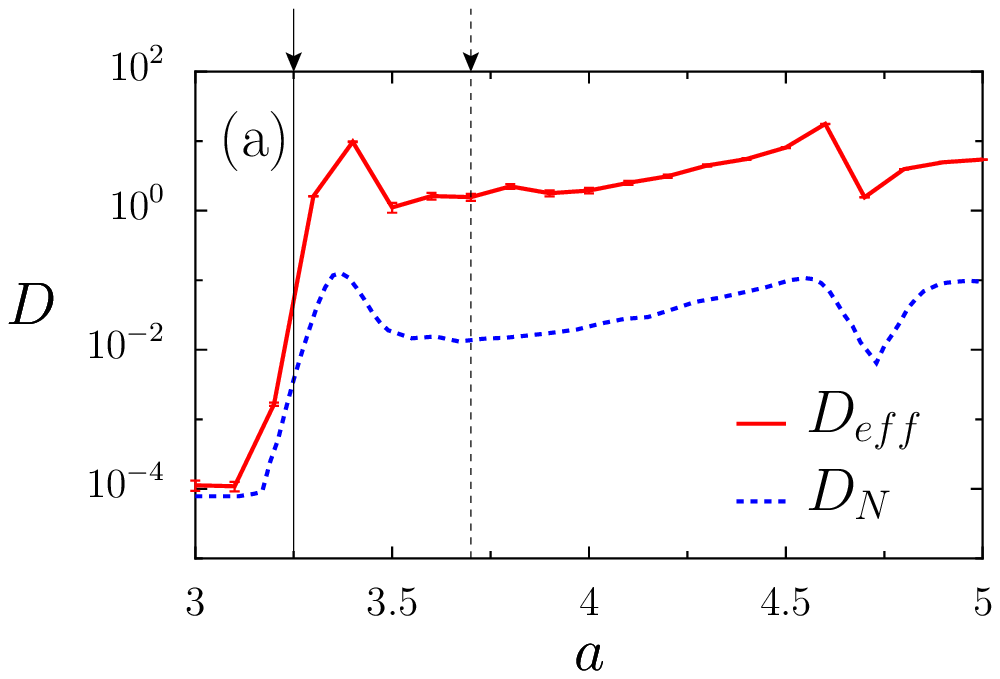}
\hfill
\includegraphics[scale=0.7]{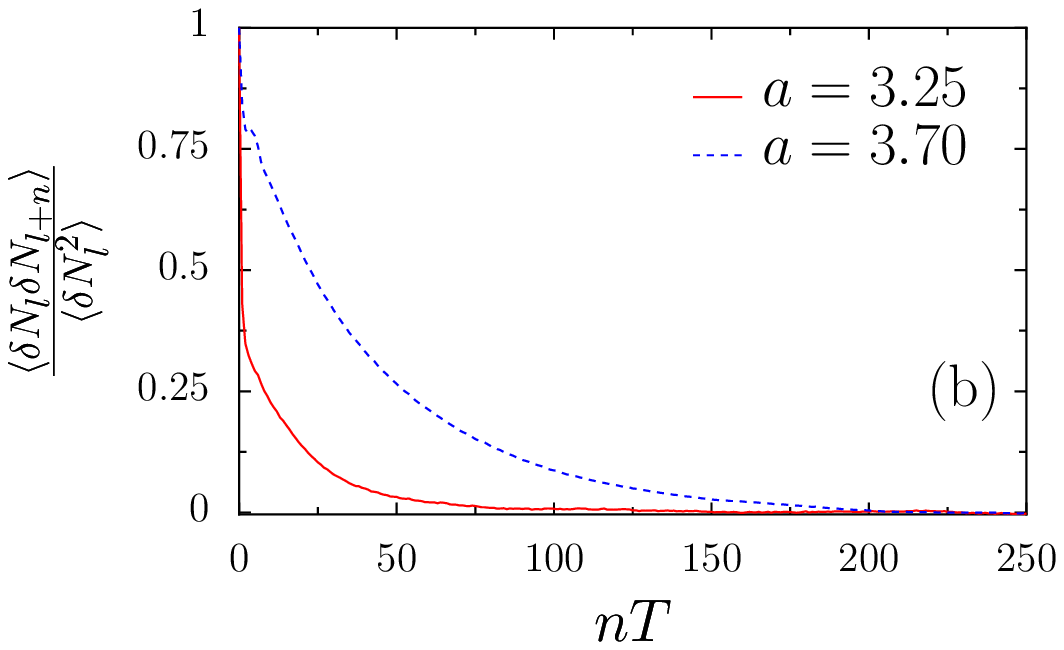}
\caption{The step number diffusion $D_N$ and the effective
diffusion $D_{eff}$ are depicted in panel (a) as  functions of the 
driving amplitude
$a$  on a logarithmic scale for the driving frequency $\omega =4.9$,
noise strength $D=0.001$ 
and potential parameters $c_1=0.425$ and $c_2=0.04$. 
In the
locked regime for $a<3.1$ both diffusion constants are comparably
small. In the running regime for $a>3.1$ both diffusion constants
are large with $D_{\text{eff}}$ becoming larger than $D_N$ roughly by a factor
of ten. The normalized correlations $\langle \delta N_{l+n} \delta
N_l\rangle/  \langle (\delta N_{l})^2 \rangle $
are depicted in panel (b) for two different values of $a$, which are marked by
arrows in panel (a).   
They extend over many periods $T$ leading to the observed
discrepancy of the two diffusion constants. For $a=3.25$ the decay of
the correlations is much faster than for $a=3.7$. Accordingly the
difference $D_{\text{eff}}- D_N$ is more pronounced at the larger $a$ value.} 
\label{f2}
\end{center}
\end{figure}

\begin{abstract}
  The expression for the effective diffusion of an inertial, periodically driven
  Brownian particle in an asymmetric, periodic  potential is compared 
  with the step number diffusion  which is extracted from the corresponding 
  coarse grained hopping process specifying the number of covered spatial 
  periods within each temporal period. 
  The two expressions are typically different and involve the correlations 
  between the number of hops.
\end{abstract}
The expression used for the diffusion constant $D_{\text{ eff}}$ in
eq. (12) in Ref.~\cite{MKMTHL}, 
which will be denoted by $D_N$ in the sequel, coincides with
the definition in eq. (5) in Ref.~\cite{MKMTHL} only under certain 
conditions, see
the Fig.~\ref{f2}.
In order to understand the relation between these expressions we split
the random distance $x(nT)$, which the particle has
covered after n periods of duration $T$, into its
integer multiple $N(nT)$ of spatial periods of length $L$ and a
remainder $\epsilon(nT)$
\begin{equation}
x(nT) = N(nT) L + \epsilon(nT)
\end{equation}
Note that $\epsilon(nT)$ is non-negative and
bounded by $L$.
From the definition (5) we then obtain
\begin{eqnarray}
D_{\text{eff}}& =& \lim_{n \to \infty}\left \{ \frac{L^2 
    \langle  (\delta N(nT))^2 \rangle }{2 n T} +
   \frac {L \langle \delta N(n T) \delta
   \epsilon(nT) \rangle }{n T} +  \frac { \langle ( \delta
   \epsilon(nT))^2 \rangle }{2 n T} \right \}\nonumber \\
&=& \lim_{n \to \infty} \frac{L^2 
    \langle  (\delta N(nT))^2 \rangle }{2 n T} 
\label{DeffErratum}
\end{eqnarray}
Each fluctuation $\delta N(nT) = N(nT) - \langle N(nT) \rangle$
contributes to the averages with a factor growing as 
$n^{1/2}$ such that
only the first term on the right hand side of the first equation
contributes in the limit $n \to \infty$. 
Next, we represent the number $N(nT)$ of spatial periods covered within $n$
temporal periods $T$ as the sum of the number $N_k$ of spatial periods
which the particle passes through within
the $k$th temporal period of the driving force, i.e.
\begin{equation}
N(nT) =    \sum_{k=1}^n N_k
\end{equation}
From eq.~(\ref{DeffErratum}) we then obtain
\begin{eqnarray}
D_{\text{eff}}& =& \lim_{n \to \infty} \frac{L^2 \sum_{k,l}^n
    \langle  \delta N_k \delta 
    N_l \rangle }{2 n T} \nonumber \\
&=& D_N +\lim_{n \to \infty}  
    \frac{L^2 \sum_{k,l,k\neq l}^n
    \langle  \delta N_k \delta 
    N_l \rangle }{2 n T}
\end{eqnarray}
where
\begin{eqnarray}
 D_N & = &\lim_{n \to \infty} \frac{L^2 \sum_k^n
    \langle  (\delta N_k)^2 \rangle }{2 n T} \nonumber \\
&=& \frac{L^2 \langle (\delta N_k)^2 \rangle}{2T}
\end{eqnarray}
is the quantity that was used in eq. (12) in Ref.~\cite{MKMTHL}.
Here we took into account that in the limit $n \to \infty$ the increments  
$\delta N_k$ become stationary and
the variances $\langle (\delta N_k)^2 \rangle$ are independent of $k$.   
The difference between $D_{\text{eff}}$ and $D_N$ results
from the sum over the correlations between the increments $\delta
N_k$. In the limit $n \to \infty$ the double sum is dominated by terms
with large values of $k$ and $l$ for which the correlations $\langle
\delta N_k \delta N_l \rangle$ only depend on the difference $k-l$. 
If the correlations decay faster than $(k-l)^{-2}$ the limit can be
simplified to read
\begin{eqnarray}
D_{\text{eff}}-D_N &=&\lim_{n \to \infty}  
    \frac{L^2 \sum_{k,l,k\neq l}^n
    \langle  \delta N_k \delta 
    N_l \rangle }{2 n T}\nonumber \\
&=& \frac{L^2}{T} \sum_{m=1}^\infty \langle  \delta N_{k+m} \delta 
    N_k \rangle 
\end{eqnarray}
In principle, this sum may take positive as well as negative values. In the
Fig.~\ref{f2} we display the dependence of
$D_{\text{eff}}$ and $D_N$ on the amplitude of
the driving force and illustrate illustrate the correlation functions 
$\langle  \delta N_{k+m} \delta N_k \rangle$ in two selected cases.


\Bibliography{99}


\bibitem{BM} H\"anggi P and Bartussek R 1996 {\it Lect. Notes Phys.} \textbf{476} p~294; 
J\"ulicher F, Ajdari A and Prost J, 1997 {\it Rev. Mod. Phys} {\bf 69} p~294;
Porto M, Urbakh M and Klafter J, 2001 {\it J. Lumin.} {\bf 94} p~137;
Astumian R D and H\"anggi P, 2002 {\it Physics Today} \textbf{55} (No.11) p~33; 
Reimann P and H\"anggi P, 2002 {\it Appl. Phys. A} \textbf{75} p~169; 
Reimann P, 2002 {\it Phys. Rep.} \textbf{361} p~57; 
Linke H, 2002 {\it Appl. Phys. A} \textbf{75} p~167; 
H\"anggi P, Marchesoni F, and Nori F, 2005 {\it Ann. Phys. (Leipzig)} {\bf 14} p~51.
%
\bibitem{jones04} Jones P H, Goonasekera M and Renzoni F, 2004 {\it Phys. Rev. Lett.} {\bf 93} 073904.
%
\bibitem{lee05} Lee Sang-Hyuk, Ladavac K, Polin M and Grier D G, 2005 {\it Phys. Rev. Lett.} {\bf 94} 110601.
%
\bibitem{babic05} Babi\v c D and Bechinger C, 2005 {\it Phys. Rev. Lett.} {\bf 94} 148303.
%
\bibitem{vondel05} Van de Vondel J, de Souza Silva C C, Zhu B Y,
Morelle M and Moshchalkov V V, 2005 {\it Phys. Rev. Lett.} {\bf 94}
057003.
%
\bibitem{villegas05} Villegas J E, Gonzalez E M, Gonzalez M P,
Anguita J V and Vincent J L, 2005 {\it Phys. Rev. B} {\bf 71}
024519.


\bibitem{peclet}
P\'eclet E, 1841 \textit{Ann. Chim Phys}  {\bf 3} p~107; P\'eclet,
E., 1843, {\it Trait\'e de la Chaleur Consider\'ee dans ses
Applications}, 3 vols., (Hachette, Paris).
%
\bibitem{jung96} Jung P, Kissner J G and H\"anggi P, 1996
{\it Phys. Rev. Lett.}  \textbf{76} p~3436.
%
\bibitem{lindner99} Lindner B,  Schimansky-Geier L, Reimann P, H\"anggi P and Nagaoka M,
1999 {\it Phys. Rev. E} {\bf 59} p~1417.
%
%
\bibitem{hanggi1982} H\"anggi P and Thomas H, 1982 {\it Phys. Rep.} {\bf 88} p~207.
%
%
\bibitem{mach} Machura L, Kostur M, Talkner P, \L uczka J, Marchesoni F and
H\"anggi P, 2004 {\it Phys. Rev. E} {\bf 70} 061105.
%
\bibitem{landau} Landau L D and Lifshitz E M, 1959 {\it Fluid Dynamics} (Oxford: Pergamon) p~203.
%
\bibitem{munakata} Suzuki D and Munakata T, 2003 {\it Phys. Rev. E} {\bf 68}
021906; Wang H and Oster G, 2002 {\it Europhys. Lett.} {\bf 57}
p~134.

%
\bibitem{talkner-rice} Talkner P, 2003 {\it Physica A} {\bf 325} p~124;
Talkner P, Machura L, Schindler M, H\"anggi P and \L uczka J, 2005
{\it New Journal of Physics} {\bf 7} 14;
Casado-Pascaul J, Gomez-Ordonez J, Morillo M, Lehmann J, Goychuk I and H\"anggi P, 2005 
{\it Phys. Rev. E} {\bf 71} 011101.
%
\bibitem{fano} Fano U, 1947 {\it Phys. Rev.} {\bf 72} p~26.
%
\bibitem{ODBM}
Bartussek R,  H\"anggi P and Kissner  J G, 1994 {\it Europhys.
Lett.} {\bf 28} p~459; 
Savel'ev S, Marchesoni F, H\"anggi P and Nori F, 2004  {\it Europhys. Lett.} {\bf  67} p~179; 
Savel'ev S, Marchesoni F, H\"anggi P and Nori F, 2004 {\it Phys. Rev. E} {\bf 70} 066109;
Savel'ev S, Marchesoni F, H\"anggi P and Nori F, 2004 {\it Eur. J. Phys. B} {\bf 40} p~403;

\bibitem{chaotic}
Mateos J L, 2001 {\it Acta Phys. Pol. B} {\bf 32} p~307; Mateos J L,
2003 {\it Physica A} {\bf 325} p~92;
Arizmendi C M, Family F and Salas-Brito A L, 2001 {\it Phys. Rev.
Lett.} {\bf 63} 061104; Larrondo H A, Family F and Arizmendi C M,
2002 {\it Physica A} {\bf 303} p~67; Larrondo H A, Arizmendi C M and
Family F, 2003 {\it Physica A} {\bf 320} p~119;
Son W-S, Kim I, Park Y-J and Kim C-M , 2003 {\it Phys. Rev. E} {\bf
68} 067201;
Barbi M and Salerno M, 2000 {\it Phys. Rev. E} {\bf 62} p~1988;
%
Borromeo M, Costantini G and Marchesoni F, 2002 {\it Phys. Rev. E}
{\bf 65} 041110; Sengupta S, Guantes R, Miret-Artes S and H\"anggi
P, 2004 {\it Physica A} {\bf  338} p~406.
%
\bibitem{giantdiff}
Schreier M Reimann P, H\"anggi P and Pollak E, 1998 {\it Europhys.
Lett.} {\bf 44} p~416; Costantini G and Marchesoni F, 1999 {\it
Europhys. Lett.} {\bf 48} p~491; Reimann P, Van den Broeck C, Linke
H, H\"anggi P, Rubi J M and Perez-Madrid A, 2001 {\it Phys. Rev.
Lett.} {\bf 87} 010602; Reimann P, Van den Broeck C, Linke H,
H\"anggi P, Rubi J M and Perez-Madrid A, 2002 {\it Phys. Rev. E}
{\bf 65} 031104.

\bibitem{porto}
Porto M, Urbakh M and Klafter J, 2000 {\it Phys. Rev. Lett.} {\bf 85} 491.




\endbib

\Bibliography{99}
\bibitem{MKMTHL}L. Machura, M. Kostur, F. Marchesoni,
P. Talkner, P. H\"anggi, J. {\L}uczka {\it J. Phys.: Condens.
Matter}  {\bf
  17} (2005) S3741-S3752. 
\endbib
\end{document}